\documentclass[journal]{IEEEtran}

\usepackage{cite}

\usepackage{amsmath,amssymb}

\usepackage{multicol}

\usepackage{graphicx}
\ifCLASSOPTIONcompsoc
 \usepackage[caption=false,font=normalsize,labelfont=sf,textfont=sf]{subfig}
\else
 \usepackage[caption=false,font=footnotesize]{subfig}
\fi

\usepackage{url}
\usepackage{color}
\usepackage{balance}

\DeclareMathOperator{\erf}{erf}

\begin{document}

\title{Statistical Static Timing Analysis of VLSI as the Statistics of Correlated Extremes}
\author{\IEEEauthorblockN{Dmytro Mishagli, Eugene Koskin, \textit{IEEE Member}, and Elena Blokhina, \textit{Senior IEEE Member}}\\
\IEEEauthorblockA{School of Electrical and Electronic Engineering,
University College Dublin, Ireland
}
}

\maketitle

\begin{abstract}
In this study, a path-based Statistical Static Timing Analysis (SSTA) is formulated as a problem within the statistics of correlated extremes. For extreme value statistics with correlations, a novel approach to studying such systems, when the correlations are small, is developed. The approach considers a system from the first principals, starting from the multivariable characteristic function. Analytical solutions to the problem of weakly correlated extremes are obtained in the form of corrections to the Gumbel distribution. These solutions are compared with Monte Carlo simulations. The applicability limits of the proposed solutions are studied. An algorithm to estimate the covariance matrix of a timing graph is proposed.
\end{abstract}


\IEEEpeerreviewmaketitle

\section{Introduction}

The timing verification of digital integrated circuits (ICs) is an absolutely essential step: scaling down analogue and digital ICs poses very strict restrictions on their operation, in particular, in terms of timing and delay. Thus, timing analysis (TA) tools are used in order to verify and optimise digital design before fabrication~\cite{b_gerez1998,r_ho01,r_visweswariah03,b_sapatnekar04,b_orshansky08,b_bhasker09,b_lavagno16}. However, with the reduction of feature size, the loss of fabricated ICs only increases, since the effect of uncertainties on the performance of circuits increases dramatically.

Process variations can be \emph{global} (also known as inter-die or die-to-die variations)~\cite{b_wolf1986,b_bhasker09,b_dietrich12,r_kahng15,r_sengupta16}. Global variations include: 
\begin{itemize}
\item Variations that appear due to changes in the temperature or the supply of chemical compounds used during the fabrication processes (deposition, etching, oxidation, and others.) 
\item Wafer-to-wafer variations in the doping level of the substrate resulting in substrate resistivity variations.
\item Die-to-die geometry variations due to misalignments of photo masks or lithography.
\end{itemize}

In addition, there are \emph{local} variations (intra-die) that may affect integrated circuits~\cite{b_orshansky08,saha10}. Local variations may appear because of random dopant fluctuations, interface roughness and random grains in the polycrystalline structure of the gate dielectric and in the metal gates.  
Local variations cause a change in individual circuit components.  

Since the size of the technology node in the previous generations of IC  was large enough, global variations were seen as a dominant source of variation in IC in the past~\cite{b_michael1993}. Local variations are becoming more and more important now because the feature size of the most recent technologies decreased so much  so it has become comparable with atomic-scale dimensions.  

A combinational circuit can be described by an equivalent \emph{timing graph}, which can be denoted as $ G = (V, E)$. In this notation, $V$ is the vertex set, and $E$ is the edge set of this graph. The vertices of a timing graph can be connected to two different types of nodes (edges), and they correspond to different types of delays arising in a combinational circuit.

\begin{figure}[!t]
\centering
\includegraphics[width=0.5\textwidth]{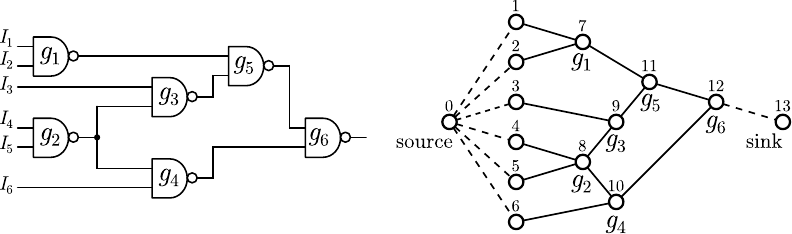}
\caption{Example of a combinational logic circuit and the construction of its timing graph with one source and sink.}
\label{fig:logic_gates_circuit_and_graph}
\end{figure}

An example of a generic combinational logic circuit and its timing graph obtained by converting the paths and gates into a set of vertices and edges is shown in Fig.~\ref{fig:logic_gates_circuit_and_graph}. In this timing graph, one can identify multiple paths from the source to the sink of the circuit. Timing analysis deals with the analysis of the timing graph and its paths. There is a simple transform that converts a default timing graph into a one with the single source and single sink.

Returning to the verification of ICs, their timing graphs and timing analysis, we recall that the nodes in a graph represent delays in a signal. The time when a signal arrives at a gate (node) in a circuit can vary due to many reasons. The input data may experience errors, the temperature and voltage supplying a logic circuit may change, and there are manufacturing differences. Hence, the main goal of \emph{Static Timing Analysis}  is to verify that despite these possible variations, all signals will arrive neither too early nor too late, and hence proper circuit operation can be assured. 

In STA, the word \emph{static} implies that this timing analysis is carried out in an input-independent manner. The purpose of this analysis is to find the worst-case delay of a given circuit considering all possible input combinations. It is known that the computational efficiency of STA is excellent, and, hence, it is a common circuit verification tools, although it has a number of limitations. In particular, STA cannot  handle intra-die variations and spatial correlation easily. If one wants to consider local variations, STA needs many corners to handle all possible cases. In addition, if there are significant random variations, STA predictions become too pessimistic. 

Known limitations of STA have led to the development of 
\emph{SSTA}, which replaces deterministic timing of gates and interconnects with probability distributions. The outcome of SSTA is a distribution of possible circuit outcomes rather than a single outcome~\cite{b_srivastava05}. There are two main approaches with the  broad context of SSTA: \emph {path-based} and \emph{block-based} methods~\cite{b_sapatnekar04,r_blaauw08}.

A path-based approach looks into the total gate and interconnect delays on specific paths. The statistical foundation of the method may look simple at a glance. This approach requires that  the paths of interest are identified prior to starting the analysis. While it sounds like a straightforward technique, a significant challenge may arise in the case of non-Gaussian correlated distributions. Another potential issue is that some other paths, which are not included in the original selection,  may be relevant but not analysed. Hence, path selection and correlation analysis are extremely important.

Closely related to the path-based SSTA are Monte Carlo (MC) methods. While MC analysis of ICs is \textit{de facto} golden standard for the designers, it has become extremely expensive. Thus, there is ongoing research aiming to reducing the simulation time while keeping MC's accuracy. One of the problems of MC-based methods is related to random samples generation~\cite{tasiran06,kanj06,veetil07,veetil11}. A quantile extraction algorithm for an MC simulation was proposed in~\cite{krishnan18}. The usage of order statistics~\cite{b_david} allowed one to decrease the number of samples required for MC simulations. In~\cite{lange14,lange15,lange16}, an approach to model variability based on generalised lambda distributions is proposed.

A block-based approach looks into the arrival times and operation times for each node (each logic gate). The advantage of this method is completeness as it does not require any path selection. The main challenge is that the statistical $\max$ and $\min$  operations, with and without the presence of correlations, is needed, and it is recognised to be a hard technical problem in Statistics. We refer the Reader to the review papers~\cite{r_nitta07,r_blaauw08,r_forzan09} for further details on the block-based methods.

In this paper, the path-based SSTA problem is formulated as that of Extreme Values Statistics. Then, the problem is relaxed to the case of weakly correlated RVs. Analytical solutions to the problem of weakly correlated RVs are obtained in the form of corrections to the Gumbel distribution. These solutions are compared against MC numerical simulations. The applicability limits of the proposed solutions are studies. The path-based approach to SSTA formulated in this study requires the covariance matrix of a timing graph, therefore, an algorithm to estimate the matrix is proposed. This research continues and extends the results of the work~\cite{mishagli19}, presented at the ISCAS 2019 (Sapporo, Japan).

The paper is organised as follows. In Section~II, we outline the basics of path-based SSTA, make necessary statistical preliminaries and present a general statement of the problem. In Sections III and IV, we develop an asymptotic theory for extreme value statistics, assuming that correlations are weak. These sections somewhat repeat the work~\cite{mishagli19}. In Section~V, the higher terms of the expansion of CDF (PDF) of correlated extremes are considered. Section~VI discusses deviation from IID case. An algorithm for estimation of the path covariance matrix is presented in Section~VII. Finally, overall discussion of the obtained results concludes this papers with Section~VIII.

\section[Statement of the Problem]{Statement of the Problem in Path-Based SSTA}
\sectionmark{Statement of the Problem}

Consider a simple combinational logic circuit and its timing graph shown in Fig.~\ref{fig:logic_gates_circuit_and_graph}. 
In its most general formulation, the problem addressed within path-based SSTA for such a circuit is to determine extreme delays arising in it. This problem 
is equivalent to the problem of finding the maximum of a set of $N$ \emph{correlated} RV:
\begin{equation}\label{chap4:eq:delay}
    \zeta = \max(X_1, X_2, \ldots, X_N),
\end{equation}
where each RV component $X_i$ describes the accumulated (total) delay of the relevant path of a circuit. 

\begin{figure}[!t]
\centering
\includegraphics[width=0.45\textwidth]{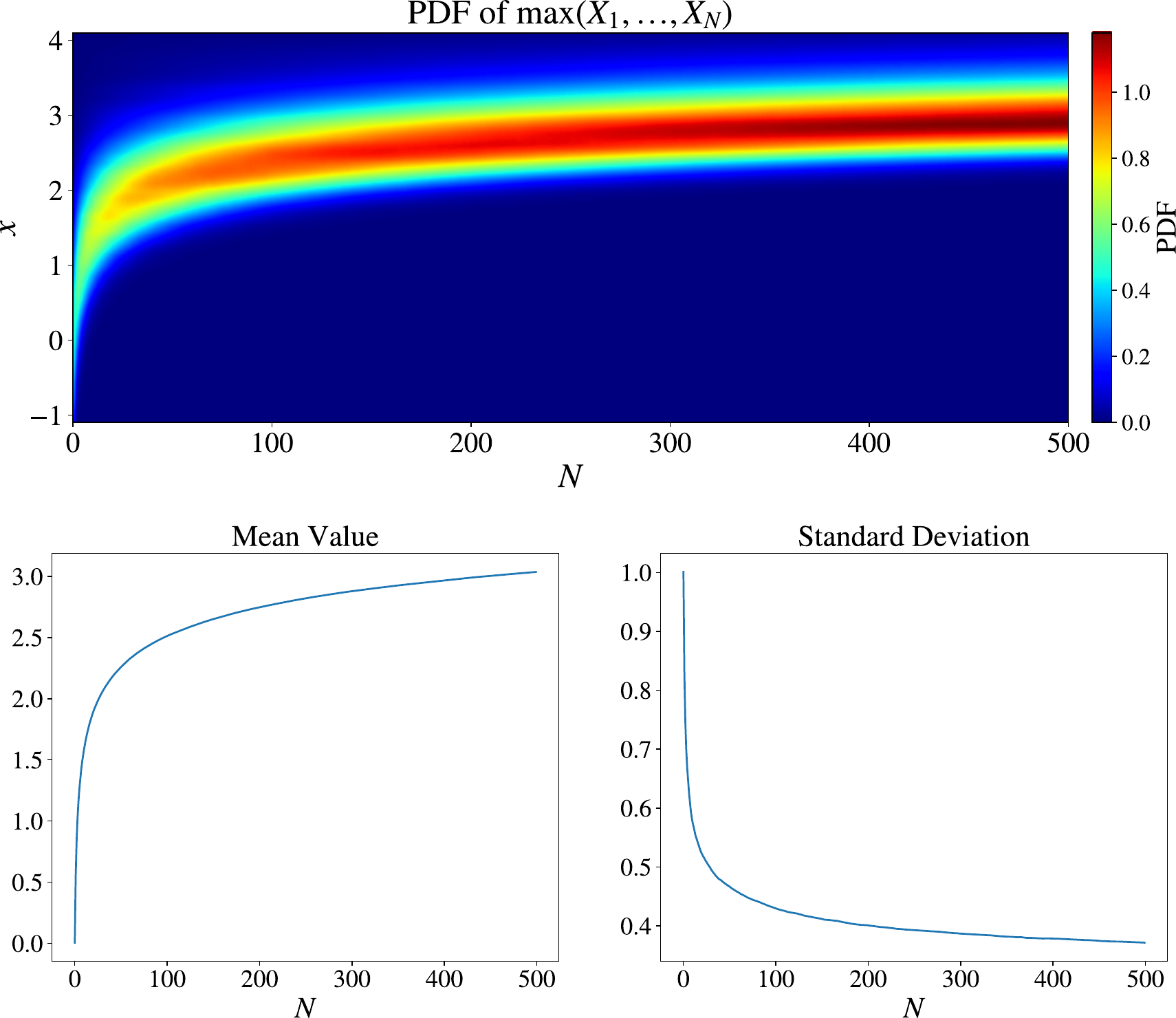}
\caption{Illustration of the statistics of extremes. \textit{On the top}: a heatmap with the probability density function of the RV $\zeta$, which is the maximum of a set of $N$ standard Gaussian IID Random Variables, $X_i \sim \mathcal N(0,1)$, against the number $N$. \textit{On the bottom}: the man value and the standard deviation for the same RV $zeta$. One can see that the mean value increases (standard deviation decreases) with the increase of the number $N$.}
\label{fig:evs_illustration}
\end{figure}

Since the number of nodes in timing graphs of  modern VLSI systems is typically of the order of $10^9$ and greater,~see Ref.~\cite{b_dietrich12,b_lavagno16},  one can conclude that the number of delay contributions in a given path is sufficiently large in order to satisfy the requirements of the Central Limit Theorem. Thus, without any loss of generality, we assume that all of the RVs $X_i$ that describe the accumulated delay along the $i^{\text{th}}$ path in a graph $G$, are distributed according to the Gaussian distribution (see Fig.~\ref{fig:evs_illustration}).

In the case of zero correlations between RVs, \emph{i.e.,} IID random variables, the distribution of \eqref{chap4:eq:delay} is known and given by the Gumbel function. Let us introduce specific notations. The CDF of the Gumbel distribution will be denoted as follows:
\begin{equation}\label{chap4:eq:gumbel}
  \Psi_N(z) = \exp\left(-e^{ - \frac{z-\alpha}{\beta}} \right),
\end{equation}
where
\begin{equation}\label{chap4:eq:scaling_const}
  \alpha = \Phi^{-1}\left( 1- \frac1N \right), \quad \beta = \frac{\sqrt{2\pi}}{N \cdot \varphi(\alpha)},
\end{equation}
and the subscript $N$ highlights the number of components in the RV vector. The function $\Phi^{-1}(\cdot)$ is the inverse of the Gaussian CDF:
\begin{equation}\label{eq:gauss_cdf_1}
  \Phi(x) \stackrel{\text{def}}{=} \frac{1}{\sqrt{2\pi}} \int\limits_{-\infty}^x \varphi(x') dx' = \frac12 \left[ 1 + \erf \left( \frac{x}{\sqrt{2}} \right) \right],
\end{equation}
where the standard definition of the error function is used:
\begin{equation}\label{eq:error_function_definition}
  \erf(x) = \frac{2}{\sqrt{\pi}} \int\limits_0^x e^{-x'^2} dx'.
\end{equation}
The function $\varphi(\cdot)$ is referred here as the Gaussian kernel, and is related to the normal distribution:
\begin{equation}\label{eq:gauss_pdf}
  \omega_0(x|\mu,\sigma) \stackrel{\text{def}}{=} \frac{1}{\sqrt{2\pi}\sigma} \varphi \left( \frac{x-\mu}{\sigma} \right), \quad \varphi(x) \stackrel{\text{def}}{=} e^{-\frac12 x^2},
\end{equation}

%

The mean $\tilde\mu_N$ and variance $\tilde\sigma^2_N$ of the Gumbel distribution are expressed in terms of the parameters $\alpha$ and $\beta$ as follows:
\begin{equation}\label{eq:gumbel_mean}
  \tilde\mu_N = \alpha + \gamma \beta, \quad \tilde\sigma^2_N = \frac{\pi^2}{6} \beta^2,
\end{equation}
where $\gamma \approx 0.5772$ is the Euler--Mascheroni constant.
The PDF $\psi_N(z)$ of the Gumbel distribution is given by the derivative of \eqref{chap4:eq:gumbel}:
\begin{equation}
    \psi_N(z) = \frac{1}{\beta} \exp\left(-e^{ - \frac{z-\alpha}{\beta}} - \frac{z-\alpha}{\beta} \right)
\end{equation}

In the case of the EVS of \emph{correlated} random variables (which is usually the case for most of practical applications, including VLSI circuits) the distribution of the extreme variable~\eqref{chap4:eq:delay} is not known. However, there are special cases that allow exact solutions, which give approximations to the statistics of variable~\eqref{chap4:eq:delay}, namely strong and weak correlation cases.

For \emph{strongly} correlated RVs, there is no known universal approach to undertake. However, there are particular cases and systems where solutions were obtained by employing different methods. For instance:
\begin{itemize}
    \item Extreme statistics in random matrix theory~\cite{tracy1994,tracy1996,r_majumdar14}. The joint PDF of the eigenvalues of a random Gaussian $N \times N$ matrix is known~\cite{b_mehta04}, and can be interpreted as a Gibbs-Boltzmann weight for a Coulomb gas~\cite{dyson1962}. The challenge was to study the fluctuations of the largest eigenvalue when $N\rightarrow\infty$. It was shown that it is given by the so-called \emph{Tracy-Widom distribution}\footnote{In \cite{biroli07}, this distribution was referred to as \it{``the most exciting recent result in mathematical physics''}.}~\cite{tracy1994,tracy1996}. One should note that the same distributions appear in quite different problems: in finance~\cite{biroli07}, growth models~\cite{prahofer00}, polymers~\cite{johansson00,baik00}, mesoscopic fluctuations in quantum dots~\cite{lemarie13}, and many others.
    \item Extreme value statistics in the presence of a global constraint. Such a constraint (\emph{e.g.}, it can be a conservation law) results in strong correlations between RVs. An example that allowed a solution is known as the zero range process~\cite{r_evans05}, which is a 1D lattice of $N$ interacting agents with periodic boundary conditions. The global constraint (and, therefore, correlations) is realised mathematically via $\delta$-functions.
\end{itemize}
The detailed discussion of these and other cases can be found in the review papers~\cite{r_fortin15,r_majumdar20}, as well as in the literature cited there.

\begin{figure}[!t]
\centering
\includegraphics[width=0.45\textwidth]{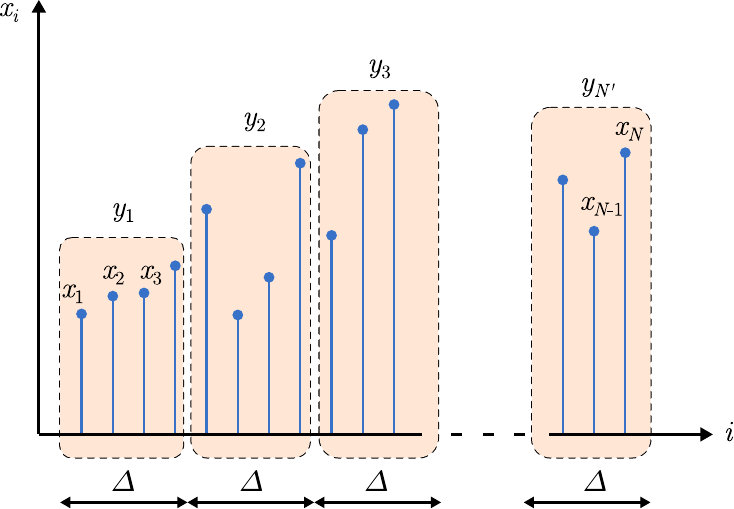}
\caption{Illustration of a set of weakly correlated RVs studied within the Renormalisation Group approach~\cite{gyorgyi08,gyorgyi10}}
\label{chap4:fig:weak_evs_illustration}
\end{figure}

The system of \emph{weakly} correlated RVs is traditionally studied via the Renormalisation Group (RG) approach~\cite{gyorgyi08,gyorgyi10}. The idea is illustrated in Fig.~\ref{chap4:fig:weak_evs_illustration} and is as follows. For a set of $N$ correlated RVs, consider \emph{blocks} in such a way, so that the RVs become non-correlated between the blocks. This is possible to achieve if the correlation between the variables decays exponentially, and is negligible after some characteristic correlation length $\Delta$:
\begin{equation}
    \langle X_iX_j \rangle - \langle X_i \rangle \langle X_j \rangle \sim e^{-\frac{|i-j|}{\Delta}}.
\end{equation}
Here, the angled brackets $\langle \cdot \rangle$ denote the averaging operation. Thus, one obtains a set of new \emph{uncorrelated} RVs, $y_1, y_2, \ldots, y_{N'}$, where $N' = N/\Delta$. Each $y_i$ represents the maximum in the corresponding block. The problem, however, lies in determining this maximum within each block, as RVs within blocks remain correlated strongly. Therefore, the methods of estimating $y_1, y_2, \ldots, y_{N'}$ should be applied depending on the system under consideration, and the problem reduces to the strongly correlated cases considered above.

In this study, for the purposes of SSTA, we consider the problem of finding the distribution of the maximum of weakly correlated Gaussian RVs. The problem is stated in the most general form: starting from the \emph{first principles}, we aim to find the expression for the PDF of correlated Gaussian RVs and  the maximum of a set of RVs. This allows us to obtain corresponding corrections to the Gumbel law~\eqref{chap4:eq:gumbel}, and this constitutes the original research contribution of this paper.

\section{Weak Correlation Approximation}

In this section, we start by deriving an expansion for the multivariate  PDF $\omega(\mathbf r)$ of $N$ weakly-correlated Gaussian RVs $X_i$. We will use this expansion in later Sections to obtain the main results of this work. Let us use the notation of the random vector $\mathbf r = (x_1, x_2, \ldots, x_N)$  whose components are the realisations of random variables $X_i$. In order to obtain the expansion for $\omega(\mathbf r)$, we will use the concept of characteristic functions.  Since we need to distinguish between many types of functions and also between correlated and uncorrelated random variables, the following symbols will be used. The functions $\omega (\mathbf r)$ and $\chi (\mathbf k)$ will denote the PDF and characteristic function of \emph{correlated} Gaussian RVs respectively, while the \emph{uncorrelated} case will be denoted using the same symbols but with the subscript `0'.

The PDF  $\omega(\mathbf r)$, in the general case, is presented as follows: 
\begin{equation}\label{eq:p.d.f._general}
  \omega(\mathbf r) = \frac{1}{(2\pi)^N} \int\limits_{E_N} \chi(\mathbf k) e^{-i \mathbf k \mathbf r} d \mathbf k \,,
\end{equation}
with the characteristic function
\begin{equation}\label{eq:characteristic_fun}
  \chi(\mathbf k) = \exp \left( i \mu_i k_i  - \frac12 \Sigma_{ij} k_i k_j\right).
\end{equation}
Here, $\Sigma_{ij}$ is the covariance  matrix such that
\begin{equation}\label{eq:cov_matrix}
  \Sigma_{ij} = \delta_{ij} \sigma_i^2 + \varepsilon_{ij}, \quad \varepsilon_{ii} = 0.
\end{equation}
In the above equations, $\mu_i$ and $\sigma_i$ are the mean and standard deviation of the RV components, which are assumed to be known. The latter allows us to factorise the exponent in \eqref{eq:characteristic_fun}:
\begin{equation}\label{eq:characteristic_fun_factorised}
  \chi(\mathbf k) = \chi_0 (\mathbf k) \cdot \exp \left( - \frac12 \varepsilon_{ij} k_i k_j \right),
\end{equation}
where $\chi_0 (\mathbf k) = \exp \left( i \mu_i k_i  - \frac12 \sigma_i^2 k_i^2\right)$ is the characteristic function of $N$ uncorrelated normal RVs.

Let us assume that the non-diagonal entries of the covariance matrix are small by absolute value, and, therefore, the characteristic function can be presented through an expansion with  the coefficients
$\varepsilon_{ij}$ acting as perturbations:
\begin{equation}\label{eq:characteristic_fun_expansion}
  \chi(\mathbf k) \bigg|_{\varepsilon_{ij}\rightarrow0} = \chi_0 (\mathbf k) + \varepsilon_{ij} \cdot \left. \frac{\partial \chi (\mathbf k)}{\partial \varepsilon_{ij}} \right|_{\varepsilon_{ij}=0} + \dots
\end{equation}
We can express now the PDF $\omega(\mathbf r)$ via the non-correlated term $\omega_0$ with the following remarkable property of the multivariate normal distribution:
\begin{equation*}
  \left. \frac{\partial \chi}{\partial \varepsilon_{ij}} \right|_{\varepsilon_{ij}=0} = \frac12 \frac{\partial^2 \chi_0}{\partial \mu_i \partial \mu_j}\,.
\end{equation*}

Indeed, limiting ourselves to the first order of smallness and substituting \eqref{eq:characteristic_fun_expansion} into \eqref{eq:p.d.f._general}, we obtain the expansion for the PDF of normal correlated random variables in the weak correlation case:
\begin{equation}\label{eq:p.d.f._expansion_a}
  \omega(\mathbf r) \simeq \omega_0(\mathbf r) +  \frac{\varepsilon_{ij}}{2} \frac{\partial^2 \omega_0(\mathbf r)}{\partial\mu_i \partial\mu_j}\,.
\end{equation}

At this stage, we can introduce the explicit form for the PDF $\omega_0(\mathbf r)$. According to the note made at the beginning of this Chapter, we assume that the CLT holds. Hence, the RVs in our case are seen as correlated Gaussian ones. We can 
write $\omega_0(\mathbf r)$ as the PDF of $N$  Gaussian RVs with the mean $\mu_k$ and variance  $\sigma_k$:
\begin{equation}\label{eq:normal_p.d.f._n}
  \omega_0 (\mathbf r) \overset{\text{def}}{=} \frac{1}{(2\pi)^{N/2}}\prod\limits_{k=1}^{n} \frac{1}{\sigma_k} \varphi \left( \frac{x_k-\mu_k}{\sigma_k} \right).
\end{equation}
In the case of only one RV, we simply have~\eqref{eq:gauss_pdf}.

In principle, the second derivative of $\omega_0 (\mathbf r)$ can be written explicitly:
\begin{equation*}
    \frac{\partial^2 \omega_0(\mathbf r)}{\partial\mu_i \partial\mu_j} =
  \frac{\omega_0 (\mathbf r)}{\sigma_i^2} \left[ \frac{(x_i-\mu_i)(x_j-\mu_j)}{\sigma_j^2} - \delta_{ij} \right],
\end{equation*}
where $\delta_{ij}$ is the Kronecker delta symbol. However, this step is not necessary for the present discussion. Instead, we will proceed to differentiation by components:
\begin{equation}\label{eq:p.d.f._expansion_b}
  \omega(\mathbf r) \simeq \omega_0(\mathbf r) + \frac{\varepsilon_{ij}}{2} \frac{\partial^2 \omega_0(\mathbf r)}{\partial x_i \partial x_j}.
\end{equation}

It is easy to see that expressions \eqref{eq:p.d.f._expansion_a} and \eqref{eq:p.d.f._expansion_b} are identical. At the same time, representation~\eqref{eq:p.d.f._expansion_b} allows one to treat the mean values, $\mu_i$ and $\mu_j$, as parameters, which simplifies the analysis below.
In the next Section, we shall consider the CDF $\mathfrak F(x)$, assuming that the RVs $X_i$ have the PDF in the form of \eqref{eq:p.d.f._expansion_b}.

\section[Statistics of Weakly Correlated EVs]{Statistics of Weakly Correlated Extreme Values: First Order Approximation}

\sectionmark{Statistics of Weakly Correlated EVs}

\subsection{Derivation of the Cumulative Distribution Function in the First Order Approximation}
\label{chap4:sec:first-order-analytical} 
To find the CDF $\mathfrak F(z)$, we recall that the probability $\mathbb{P}$ that all of the RVs $X_1,X_2,\ldots,X_N$ are less than some given number $z$, $\mathbb P[\zeta<z]$, is
\begin{equation}\label{eq:c.d.f._max_general}
  \mathbb{P} [\zeta < z] \equiv \mathfrak F(z) = \int\limits_{-\infty}^{z} \ldots \int\limits_{-\infty}^{z} \omega(\mathbf r) d\mathbf r,
\end{equation}
where $d\mathbf r = dx_1dx_2 \ldots dx_N$ is the elementary volume in the $N$-dimensional space of random variables $X_i$.

We have not imposed any restrictions on the mean values and standard deviations, $\mu_i$ and $\sigma_i$, while deriving expansion \eqref{eq:p.d.f._expansion_b}. For the sake of simplicity, we assume that $\mu_i$ and $\sigma_i$ are the same for all $X_i$, \emph{i.e.,} we consider a set of IID random variables. In such a case, RVs can be re-scaled to be standard normal ones, $X_i \sim \mathcal N (0,1)$, using the transform:
\begin{equation*}
    X_i \rightarrow \frac{X_i-\mu}{\sigma}.
\end{equation*}
The effect of a deviations from the IID case is discussed below in the next Section.

Substituting expansion \eqref{eq:p.d.f._expansion_b} into \eqref{eq:c.d.f._max_general} and letting $N\rightarrow\infty$, we get two terms. The first term leads to the CDF of the Gumbel distribution:
\begin{displaymath}
  \int\limits_{-\infty}^z \ldots \int\limits_{-\infty}^z \omega_0(\mathbf r) d\mathbf r = \Phi(z)^N \underset{N\rightarrow\infty}{=} \Psi_N (z),
\end{displaymath}
while the second one involves the integral of the form
\begin{displaymath}
  I = \int\limits_{-\infty}^z \ldots \int\limits_{-\infty}^z \frac{\partial^2 \omega_0(\mathbf r)}{\partial x_i \partial x_j} d\mathbf r.
\end{displaymath}
The latter gives
\begin{equation*}
  \begin{aligned}
    I &= \frac{1}{\sqrt{2\pi}} \int\limits_{-\infty}^z \frac{\partial}{\partial x_i} e^{-\frac{x_i^2}{2}} dx_i \cdot  \frac{1}{\sqrt{2\pi}} \int\limits_{-\infty}^z \frac{\partial}{\partial x_j} e^{-\frac{x_j^2}{2}} dx_j
    \\
    &\times \frac{1}{(2\pi)^{(N-2)/2}} \int\limits_{-\infty}^z \ldots \int\limits_{-\infty}^z \exp \left( - \frac12 \sum\limits_{k \neq i,j} x_k^2 \right) \underbrace{dx_1 \ldots dx_N}_{\text{except } dx_idx_j}
    \\
    &\underset{N\rightarrow\infty}{=} \frac{\varphi(z)^2}{2\pi} \Psi_{N-2}(z).
    \end{aligned}
\end{equation*}
Finally, bringing all terms together and taking into account that $\Psi_{N-2}(x) \approx \Psi_N(x)$, we can write down the asymptotic expression for the CDF of the maximum of $N$ weakly--correlated Gaussian random variables in the first order approximation:
\begin{equation}\label{eq:gumbel_perturbed}
  \mathfrak F_N(z) = \Psi_N (z) \left[ 1 + \frac{\varphi(z)^2}{4\pi} \sum_{\substack{i,j\\i\neq j}} \varepsilon_{ij} + \ldots \right].
\end{equation}

The PDF of the extended Gumbel law~\eqref{eq:gumbel_perturbed} can be calculated explicitly. The generalised formula for the PDF is given in the next Section by expression~\eqref{eq:gumbel_pdf_perturbed} since it serves both cases, the first order and the second order approximations. 

Note that this formula is derived assuming the diagonal entries $\varepsilon_{ii} = 0$, and is valid for small correlations:
\begin{equation}\label{eq:smallness_condition}
  |\varepsilon_{ij}| \ll \delta_{ij} \sigma_i^2.
\end{equation}

Formula~\eqref{eq:gumbel_perturbed} represents an extension of the  Gumbel distribution and constitutes the original result of this Section. Modelling and simulation of this case are presented in the next Section. 

\subsection{Modelling and Comparison with Monte Carlo Simulations Using the First Order Approximation}\label{chap4:sec:sim_1}

To understand the result and verify its validity,  we will compare it with Monte Carlo simulations. This type of simulations is the reference method to verify a VLSI design~\cite{b_sapatnekar04}. For the sake of illustration of the obtained results, we choose the correlation between the RVs $X_i$ as follows:
\begin{equation}\label{eq:simulation_conditions}
  \langle X_i X_j \rangle = \sigma^2 \rho^{|i-j|}, \quad \langle Y_i Y_j \rangle = \delta_{ij}, \quad \langle X_i Y_j \rangle = 0,
\end{equation}
where $X_i \sim \mathcal N(0,\sigma)$ and $Y_i \sim \mathcal N(0,1)$. In this case, random samples can be generated via the simple rule:
\begin{equation}\label{eq:simulation_law}
  X_{i+1} = \rho X_i + \sigma \sqrt{ 1- \rho^2 } Y_i,
\end{equation}
where $\rho$ is the correlation coefficient. A set of $N$ samples has been generated according to rule \eqref{eq:simulation_law} with the starting point $X_0$ chosen randomly from $\sim \mathcal N(0,\sigma)$, then $\max(X_i,\ldots,X_N)$ was computed. The procedure was repeated $10^4$ times for each numerical experiment.

\begin{figure}[!t]
\centering
\includegraphics[width=0.5\textwidth]{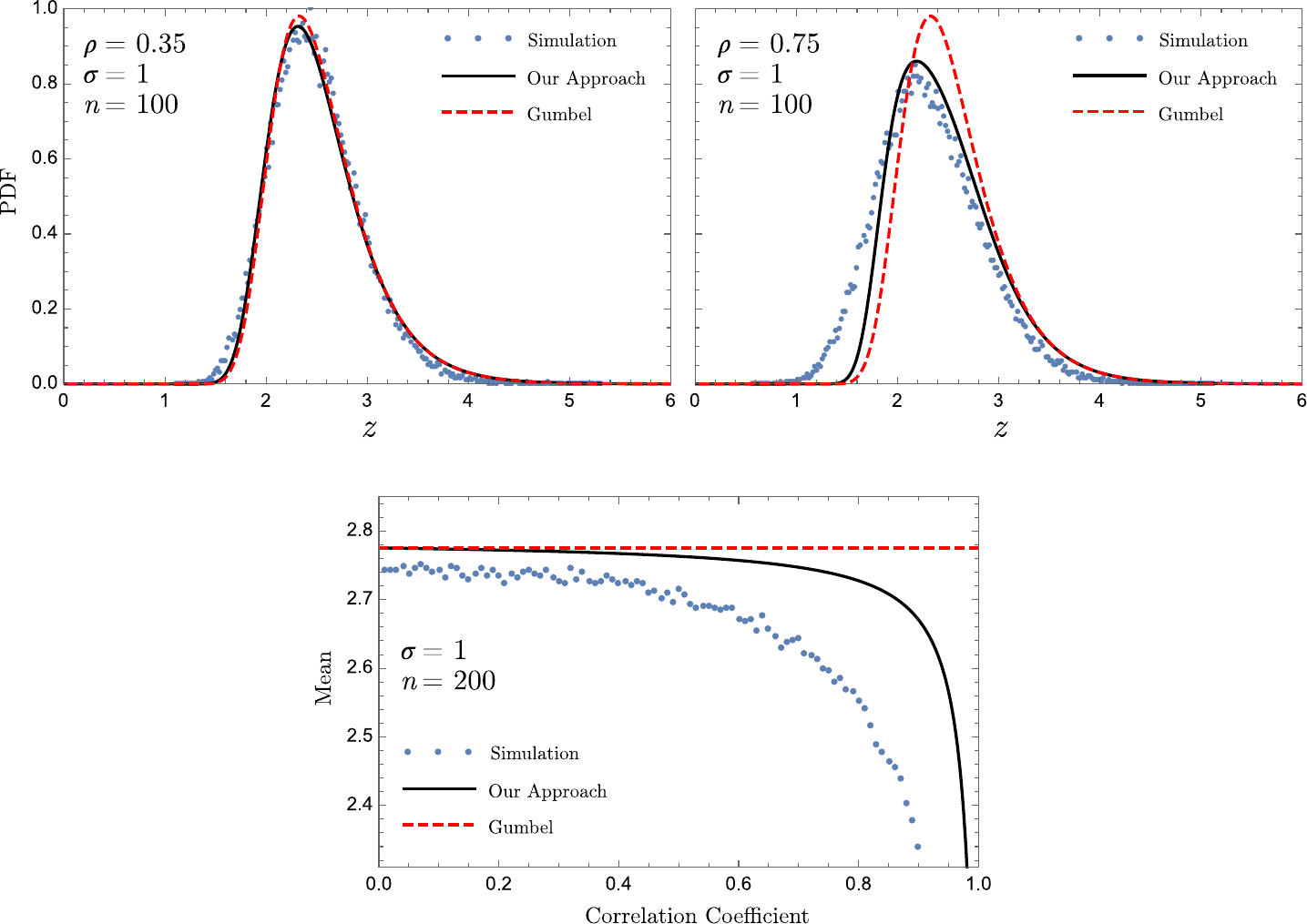}
\caption{(These results were first presented in~\cite{mishagli19}, at the ISCAS 2019, held in Sapporo, Japan.) Illustration of the proposed first-order approximation compared against the standard Gumbel distribution and Monte Carlo simulations. The upper row shows the PDF of the weakly correlated Gaussian RVs~\eqref{eq:simulation_conditions} obtain through the proposed approach and compared with the Gumbel distribution and MC simulations. Two different examples of correlations $\rho = 0.35$ and $\rho = 0.75$ are shown for the number of random variables $N=100$. The lower graph shows the behaviour of the mean as a function of the correlation coefficient $\rho$ for $N=200$. }
\label{chap4:fig:gumbel_iscas}
\end{figure}

For a large number of RVs ($N \gtrsim 10^2$), the dependence of the PDF on the correlation coefficient $\rho$ is studied. As one can see from Fig.~\ref{chap4:fig:gumbel_iscas}, for relatively small values of $\rho$, the deviation of both the perturbed and uncorrelated Gumbel distributions, \eqref{eq:gumbel_perturbed} and \eqref{chap4:eq:gumbel} respectively, from MC simulations is negligible (error in the mean value is less than 2$\%$). This implies that small correlations can be ignored without any noticeable loss of accuracy. At the same time, the comparison in the strong correlation regime, $\rho>0.5$, shows that approximation \eqref{eq:gumbel_perturbed}, even though it should not be applied for such $\rho$, is able to describe the correct trend.

Since this approach is developed with delay propagation in a logic circuit in mind, it would be particularly interesting to discuss the behaviour of the mean value as a function of the correlation coefficient $\rho$, as also shown in Fig.~\ref{chap4:fig:gumbel_iscas}.
One can see that the uncorrelated Gumbel distribution~\eqref{chap4:eq:gumbel} leads to significant deviations from the numerical simulations for $\rho>0.5$. Thus, the expression for the mean~\eqref{eq:gumbel_mean} can be interpreted as the worst case delay. At the same time, weakly-correlated formula~\eqref{eq:gumbel_perturbed} gives a result between that and the simulation dots. The deviation of \eqref{eq:gumbel_perturbed} from the MC simulations may be reduced by taking into account further terms in the expansion \eqref{eq:characteristic_fun_expansion}, as shown in the next Section.

It should be pointed out that the mean is decreasing with the increase of the correlation coefficient, which is seen both from the simulation and the theory in Fig.~\ref{chap4:fig:gumbel_iscas}. Recalling the meaning of the RV $\zeta$ in SSTA, we see that the total delay in a logic circuit is inversely proportional to the value of the correlation coefficient $\rho$. 

\section{Statistics of Weakly Correlated Extreme Values: Higher Terms}
\sectionmark{Statistics of Weakly Correlated EVs}

\subsection{Derivation of the Cumulative Distribution Function in the Second Order Approximation}

In this Section, we will try to improve the previous result by taking into account the second order term in the expansion~\eqref{eq:characteristic_fun_expansion}:
\begin{align}\label{eq:characteristic_fun_expansion_second}
  \chi(\mathbf k) \bigg|_{\varepsilon_{ij}\rightarrow0} = \chi_0 (\mathbf k) &+ \varepsilon_{ij} \cdot \left. \frac{\partial \chi (\mathbf k)}{\partial \varepsilon_{ij}} \right|_{\varepsilon_{ij}=0} \nonumber
  \\
  &+ \frac12 \varepsilon_{ij} \varepsilon_{kl} \cdot \left. \frac{\partial^2 \chi (\mathbf k)}{\partial \varepsilon_{ij} \partial \varepsilon_{kl}} \right|_{\varepsilon_{ijkl}=0}
  + \dots
\end{align}
In the similar manner as was done for the first derivative, one can show that:
\begin{equation*}
  \left. \frac{\partial^2 \chi}{\partial \varepsilon_{ij} \partial \varepsilon_{kl}} \right|_{\varepsilon_{ijkl}=0} = \frac14 \frac{\partial^2 \chi_0}{\partial \mu_i \partial \mu_j \partial \mu_k \partial \mu_l},
\end{equation*}
and present the expansion of the PDF of  correlated Gaussian RVs in the second order approximation as follows:
\begin{equation}\label{eq:pdf_expansion_second}
  \omega(\mathbf r) \simeq \omega_0(\mathbf r) + \frac{\varepsilon_{ij}}{2} \frac{\partial^2 \omega_0(\mathbf r)}{\partial x_i \partial x_j}
  + \frac{\varepsilon_{ij}\varepsilon_{kl}}{8} \frac{\partial^4 \omega_0(\mathbf r)}{\partial x_i \partial x_j \partial x_k \partial x_l}.
\end{equation}

\begin{figure}[t!]
\centering
\includegraphics[width=0.5\textwidth]{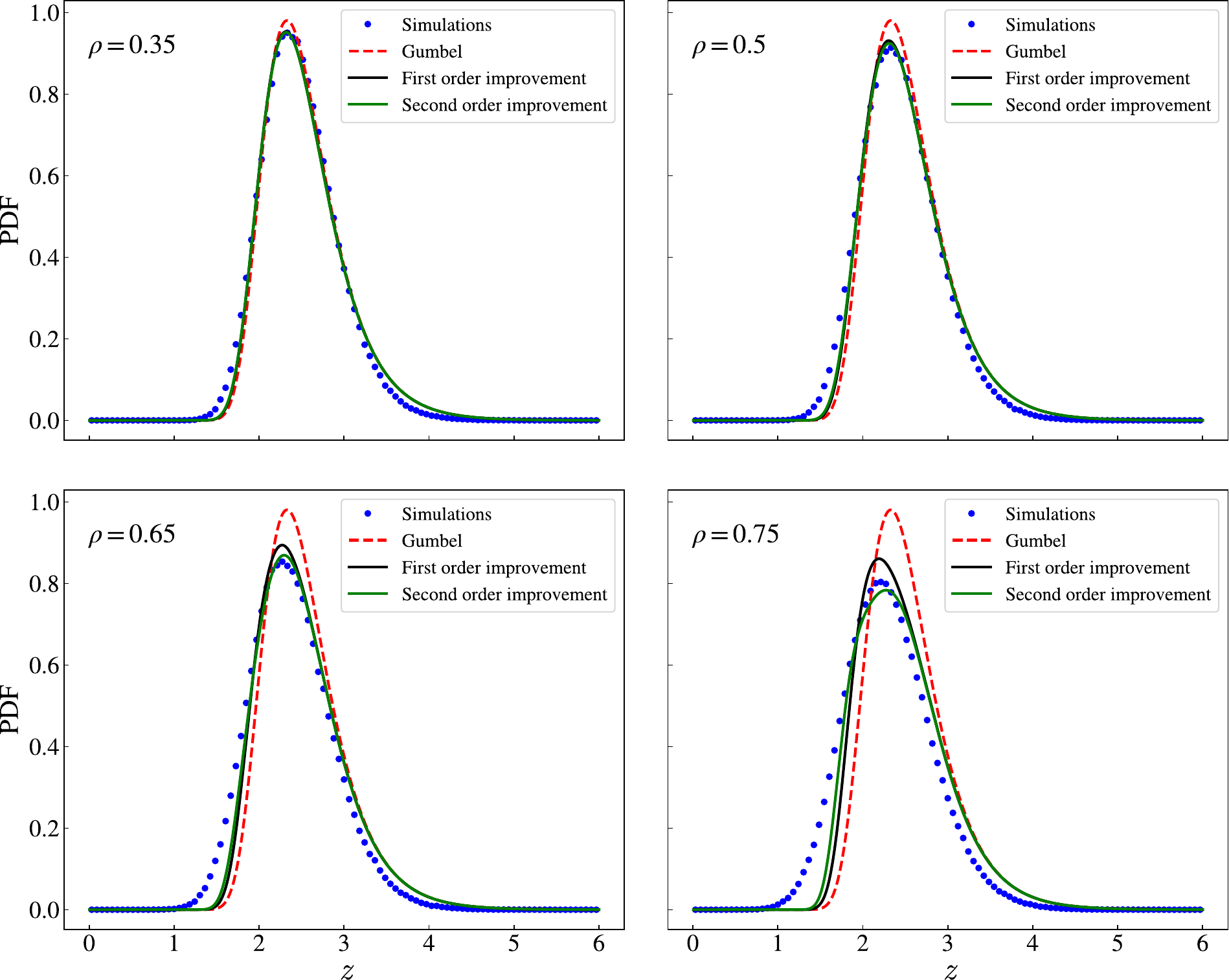}
\caption{Illustration of the proposed second-order approximation compared against the first-order approximation, standard Gumbel distribution and Monte Carlo simulations. The PDF is shown for four different correlation coefficient $\rho = 0.35$, $0.5$, $0.65$ and $0.75$, demonstrating the transition from weakly-correlated to the strongly-correlated case. The number of variables is $N=100$.}
\label{chap4:fig:gumbel_pdf_table}
\end{figure}

Now, substituting \eqref{eq:pdf_expansion_second} into \eqref{eq:c.d.f._max_general} and letting $N\rightarrow\infty$, we get the term that involves integral of the following form:
\begin{displaymath}
  I = \int\limits_{-\infty}^z \ldots \int\limits_{-\infty}^z 
  \frac{\partial^4 \omega_0(\mathbf r)}{\partial x_i \partial x_j \partial x_k \partial x_l}
  \underset{N\rightarrow\infty}{=} \frac{\varphi(z)^4}{(2\pi)^2} \Psi_{N-4}(z).
\end{displaymath}
Therefore, the asymptotic expression for the CDF of the maximum of $N$ weakly-correlated Gaussian random variables in the second order approximation has the form:
\begin{align}\label{eq:gumbel_perturbed_second}
    \mathfrak F_N(z) = \Psi_N (z) \bigg[ 1 &+ \frac{\varphi(z)^2}{4\pi} \sum_{\substack{i,j\\i\neq j}} \varepsilon_{ij} \nonumber
    \\
    &+ \frac{\varphi(z)^4}{32\pi^2} \sum_{\substack{i,j,k,l\\i\neq j,k\neq l}} \varepsilon_{ij}\varepsilon_{kl}
    + \ldots \bigg].
\end{align}
The corresponding expansion for the PDF $\mathfrak f_N(z)$ is obtained by taking the derivative of \eqref{eq:gumbel_perturbed_second}. We have
\begin{align}\label{eq:gumbel_pdf_perturbed}
    \mathfrak f_N(z) = \psi_N (z) &+ \frac{\varphi(z)^2}{4\pi} a_N(z) \sum_{\substack{i,j\\i\neq j}} \varepsilon_{ij} \nonumber
    \\
    &+ \frac{\varphi(z)^4}{32\pi^2} b_N(z) \sum_{\substack{i,j,k,l\\i\neq j,k\neq l}} \varepsilon_{ij}\varepsilon_{kl}
    + \ldots,
\end{align}
where the coefficients $a_N(z)$ and $b_N(z)$ read
\begin{equation}\label{chap4:eq:gumbel_pdf_perturbed2}
    \begin{aligned}
        a_N(z) &= \psi_N(z) - 2z\varphi(z) \Psi_n(z),
        \\
        b_N(z) &= \psi_N(z) - 12z\varphi(z)^3 \Psi_n(z).
    \end{aligned}
\end{equation}

We note that in the PDF~\eqref{eq:gumbel_pdf_perturbed}-\eqref{chap4:eq:gumbel_pdf_perturbed2}, the first perturbation term $a_N$ corresponds to the first order approximation considered in Section~\ref{chap4:sec:first-order-analytical}. The second perturbation term $b_N$ appears due to the second order approximation considered in this Section.

One can continue the expansion~\eqref{eq:gumbel_perturbed_second}, and, letting $N\rightarrow\infty$, we have
\begin{align}\label{eq:extreme_cdf_complete}
    \mathfrak F_N(z)& = \Psi_N (z) \sum_{n=0}^{\infty} \frac{1}{n!} \left( \frac{\varphi(z)^2S}{4\pi} \right)^n \nonumber
    \\
    &= \Psi_N (z) \exp \left( \frac{\varphi(z)^2S}{4\pi} \right),
\end{align}
where $S=\sum_{\substack{i,j\\i\neq j}} \varepsilon_{ij}$. Correspondingly, the PDF is given by the derivative of \eqref{eq:extreme_cdf_complete}.

Let us now examine the validity of the results obtained in this section and compare them with the first order approximation~\eqref{eq:gumbel_perturbed}.

\begin{figure}[t!]
\centering
\includegraphics[width=0.5\textwidth]{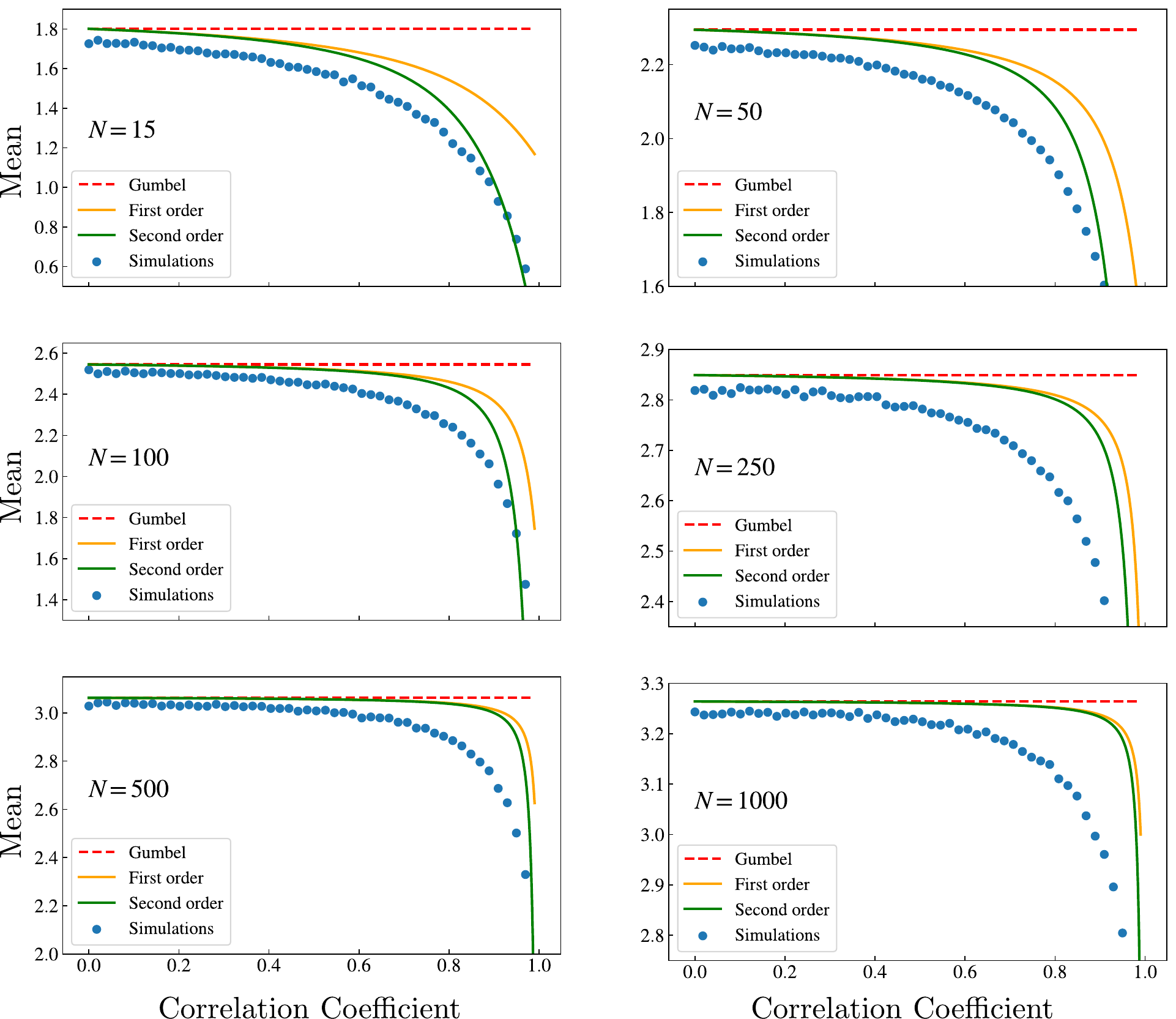}
\caption{Illustration of the proposed second-order approximation compared against the first-order approximation, the standard Gumbel distribution and Monte Carlo simulations. The mean value as a function of the correlation coefficient $\rho$ is presented.}
\label{chap4:fig:gumbel_corr_table}
\end{figure}

\subsection{Modelling and Comparison with Monte Carlo Simulations Using the Second Order Approximation} 

The verification of the second order approximation has been carried out using the same methodology outlined in Section~\ref{chap4:sec:sim_1}. Modelling and comparison with numerical simulations are summarised in Fig.~\ref{chap4:fig:gumbel_pdf_table}. This figure illustrates the proposed second-order approximation in action compared against the first-order approximation, standard Gumbel distribution and Monte Carlo simulations. The PDF of these four cases are shown for four different correlation coefficient $\rho = 0.35$, $0.5$, $0.65$ and $0.75$, demonstrating the transition from weakly-correlated to strongly-correlated case. In all presented cases, one sees a very good agreement between the PDF obtained from the second-order approximation and the MC simulation.

The investigation of the dependence of the mean value on the correlation coefficient $\rho$ is presented in Figure~\ref{chap4:fig:gumbel_corr_table}. The second order approximation seems to be particularly effective for the intermediate number $N$ of the components $X_i$. In all cases, both the first order and second order approximations can capture the correct trend even in the presence of strong correlations.

\begin{figure}[!t]
\centering
\includegraphics[width=0.5\textwidth]{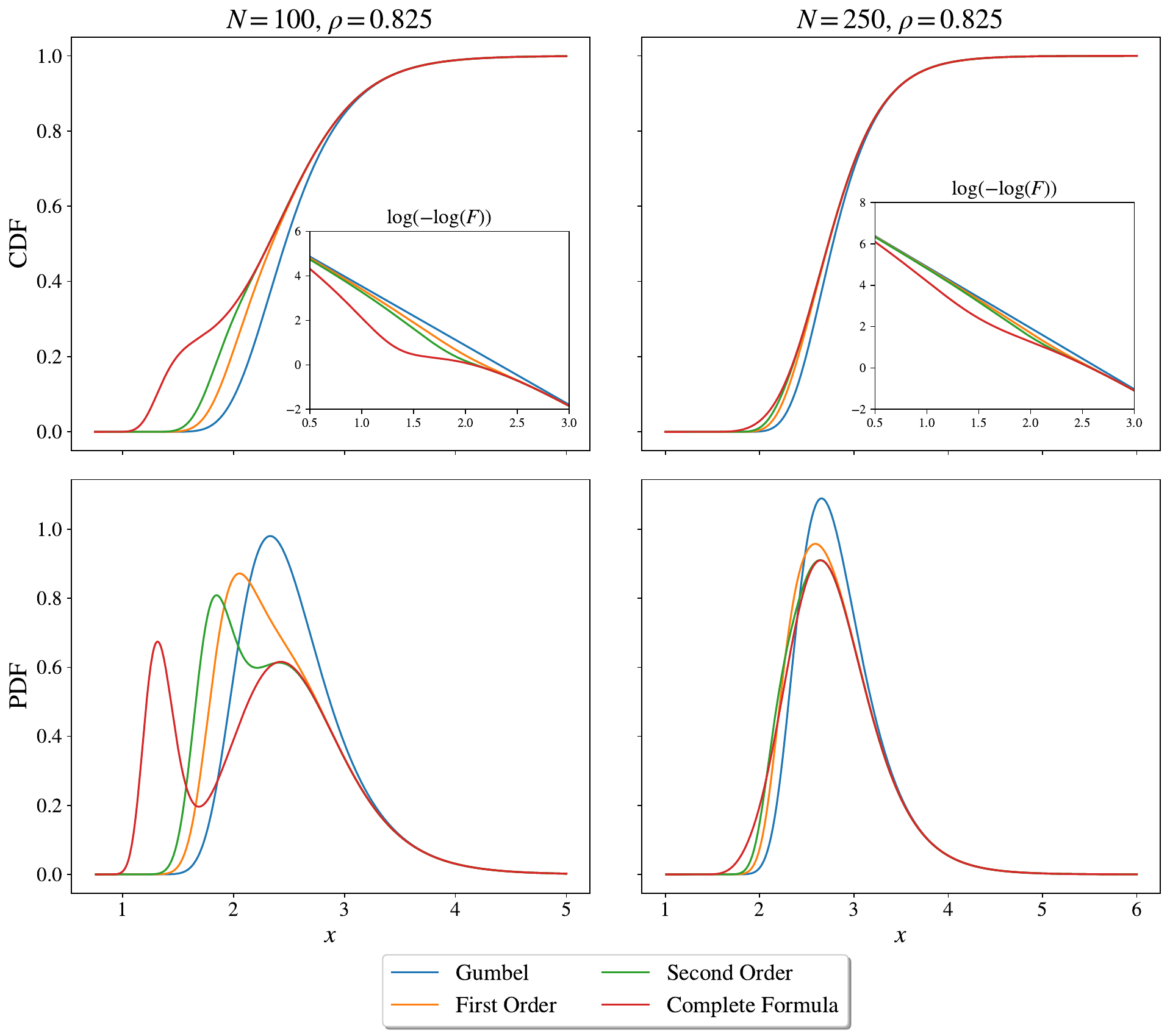}
\caption{The obtained extensions to the Gumbel law, namely, the $1^{\text{st}}$ and $2^{\text{nd}}$ approximations, and the complete formula~\eqref{eq:extreme_cdf_complete} for a case of $\rho=0.825$, where the weak correlation assumption is clearly violated. Left column shows the case with $N=100$, and one can see how the proposed corrections break down. Right column shows the case with $N=250$, and the higher number of terms in the RV~\eqref{chap4:eq:delay} leads to better performance, as role of pure Gumbel dominates. The insets show $\log(-\log(\cdot))$ of the CDF.}
\label{fig:extreme_cdf_pdf}
\end{figure}

However, the obtained asymptotic results can quickly become inapplicable for higher correlation coefficients. Hence Fig.~\ref{fig:extreme_cdf_pdf} shows Gumbel CDF \eqref{chap4:eq:gumbel} and the corrections \eqref{eq:gumbel_perturbed}, \eqref{eq:gumbel_perturbed_second}, and \eqref{eq:extreme_cdf_complete}, as well as corresponding PDFs, for the correlation coefficient $\rho=0.825$. There are two cases shown: relatively small number of variables $N=100$ (left column), and the higher number of $N=250$ (right column). For the first case, the asymptotic solutions blow (compare with Fig.~\ref{chap4:fig:gumbel_pdf_table}, where the PDFs are shown for $\rho \leq 0.75$ for the same number of variables $N=100$). For the second case ($N=250$), the corrections give good improvement of the Gumbel law. Such a behaviour is expected, as the corrections are derived assuming $N \rightarrow \infty$. Therefore,
the applicability of the results to the realistic graphs with smaller number of paths should be studied in a detail.

\section[Non-Identical Random Variables]{The Case of Non-Identical Independent Random Variables}
\sectionmark{Non-Identical Random Variables}

The major approximation we made in the previous Sections on random variables was to assume that they are identically distributed. While this assumption simplified the analysis, in reality, one does not expect that the RVs will be identical. If this approach is used to describe path delays, it not necessarily the case that all nodes in logic circuits and delays accumulated in sub-paths  will have the same mean and variance. However, we highlight again that due to the CLT, we still expect that the RVs will be Gaussian.

\begin{figure}[!t]
\centering
\includegraphics[width=0.5\textwidth]{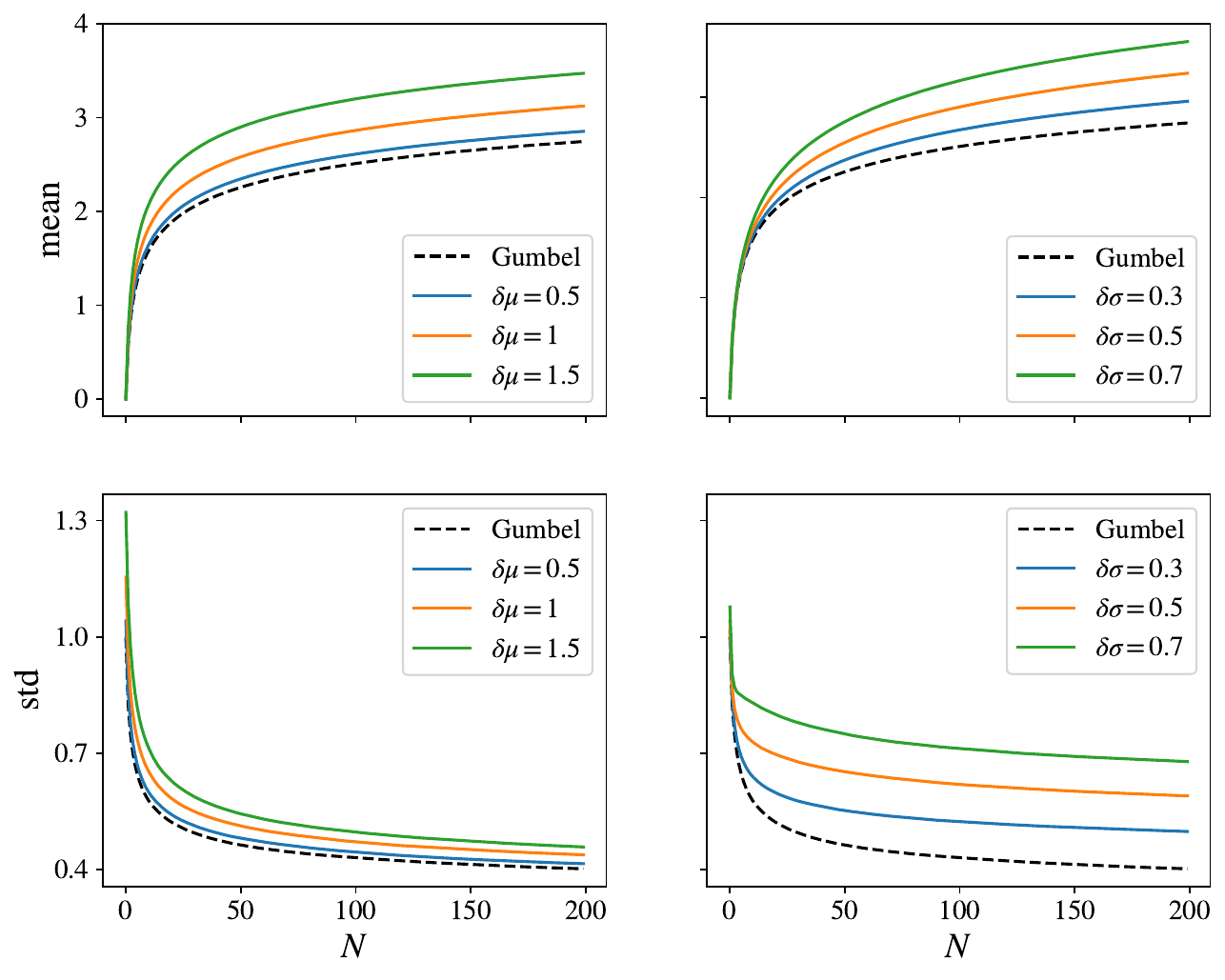}
\caption{Effect of non-identical mean values and variances (standard deviations) on the statistics of extreme values in the path-based approach. The graphs shown on the left present the dependence of the mean and standard deviation on the number $N$ of the random components $X_i$ in $\zeta$ when one varies the strength of deviations $\delta \mu$ in the mean value.  The graphs on the right show their counterparts in the case when one varies the strength of deviations $\delta \sigma$ in the standard deviation. }
\label{chap4:fig:noisy_mus_and_stds}
\end{figure}

To investigate the effect of non-identical mean values and variances in the random components $X_i$ comprising the variable $\zeta$, we consider small deviations from IID. Namely, the mean values are assumed to be $\mu_i = \mu+\xi\cdot\delta\mu$ while the variances are $\sigma_i = \sigma + \xi\cdot\delta\sigma$. Here, the variable $\xi$ is another RV drawn from the uniform distribution to help generate random deviations, \emph{i.e.}, $\xi\sim\mathcal U (-1,1)$. The quantities $\delta \mu$ and $\delta \sigma$ give the \emph{strength} of  deviations from the nominal $\mu$ and $\sigma$. For simplicity, we assume that these non-identical RVs are \emph{uncorrelated} or independent. 

The effect of non-identical mean values and variances (standard deviations) on the statistics of extreme values in the path-based approach is summarised in Fig.~\ref{chap4:fig:noisy_mus_and_stds}. The graphs shown on the left present the dependence of the mean and standard deviation on the number $N$ of the random components $X_i$ in the variable $\zeta$ when one varies the strength of deviations $\delta \mu$ in the mean value. The graphs on the right show their counterparts in the case when one varies the strength of deviations $\delta \sigma$ in the standard deviation. As a reference, the mean and standard deviation~\eqref{chap4:eq:gumbel} calculated from the Gumbel distribution of IID variables is shown by the black dashed line. The only effect one can see in this case consists in the scaling of the curves, keeping the overall shape of the Gumbel distribution. We conclude that, if required, non-identical RVs can be accommodated in the approach proposed in this paper by proper renormalisation of the relevant quantities.

\section{Path Covariance Matrix}\label{sec:cov_matrix}

The final question we would like to cover in this Chapter is related to the covariance matrix. Indeed, inspecting the example of the timing graph shown in Fig.~\ref{fig:logic_gates_circuit_and_graph}, we note that many paths starting at the Source and ending at the Sink of the graph share common nodes. This is exactly the reason why correlations appear in the random components $X_i$ in the path-based approach. Our task is to propose a method to estimate the covariance  matrix of a timing graph. 

We will assume that the delays $\tau_{ij}$ of the nodes in a timing graph are given by IID RVs drawn from a normal distribution, \emph{i.e.}, $\tau_{ij}\sim N(\mu,\sigma^2)$, where $\mu$ is the mean value of a node delay and $\sigma$ is its standard deviation. In this case, the delay between the $i^{\text{th}}$ and $j^{\text{th}}$ nodes along a path can be written as
\begin{equation}\label{eq:path_2}
  \tau_{ij} = \mu + \sigma \xi_{ij}, \quad \xi_{ij} \sim \mathcal N(0,1).
\end{equation}
Here, $\xi$ is an auxiliary random number drawn from the standard normal distribution (with a zero mean and a unit variance) to help generate a set of random delays. Thus,
\begin{equation}\label{eq:path_delay_random_xi}
  \langle \xi_{ij} \rangle = 0, \quad \langle \xi_{ij}^2 \rangle = 1, \quad \langle \xi_{ij}\xi_{kl} \rangle = \delta_{ik}\delta_{jl}.
\end{equation}

\begin{figure}[!t]
\centering
\includegraphics[width=0.33\textwidth]{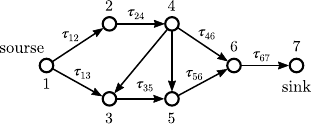}
\caption{An example of a directed acyclic graph illustrating shared nodes.}
\vspace{0.5cm}
\label{chap4:fig:graph_covmatrix_algo}
\end{figure}

Such a representation allows one to estimate the covariance matrix due to shared paths. For the sake of illustration, consider a simple directed acyclic timing graph  shown in Fig.~\ref{chap4:fig:graph_covmatrix_algo}. It is easy to see that there are four paths with the accumulated delays as follows:
\begin{equation*}
    \begin{aligned}
      &\text{path 1:} & \tau_{12} + \tau_{24} + \tau_{46} + \tau_{67} &= 4\mu + \sqrt4 \sigma \epsilon_1\\
      &\text{path 2:} & \tau_{12} + \tau_{24} + \tau_{43} + \tau_{35} + \tau_{56} + \tau_{67} &= 6\mu + \sqrt6 \sigma \epsilon_2\\
      &\text{path 3:} & \tau_{12} + \tau_{24} + \tau_{45} + \tau_{56} + \tau_{67} &= 5\mu + \sqrt5 \sigma \epsilon_3\\
      &\text{path 4:} & \tau_{13} + \tau_{35} + \tau_{56} + \tau_{67} &= 4\mu + \sqrt4 \sigma \epsilon_4\\
    \end{aligned}
\end{equation*}
The quantities $\epsilon_i$ ($i=1,\ldots,4$) describe the random part of the path delays:
\begin{equation}
    \begin{aligned}
      \epsilon_1 &= \frac12 ( \xi_{12} + \xi_{24} + \xi_{46} + \xi_{67} ) \\
      \epsilon_2 &= \frac{1}{\sqrt6} ( \xi_{12} + \xi_{24} + \xi_{43} + \xi_{35} + \xi_{56} + \xi_{67} ) \\
      \epsilon_3 &= \frac{1}{\sqrt5} ( \xi_{12} + \xi_{24} + \xi_{45} + \xi_{56} + \xi_{67} ) \\
      \epsilon_4 &= \frac12 ( \xi_{13} + \xi_{35} + \xi_{56} + \xi_{67} )\\
    \end{aligned}
\end{equation}

\begin{figure}[!t]
\centering
\includegraphics[width=0.35\textwidth]{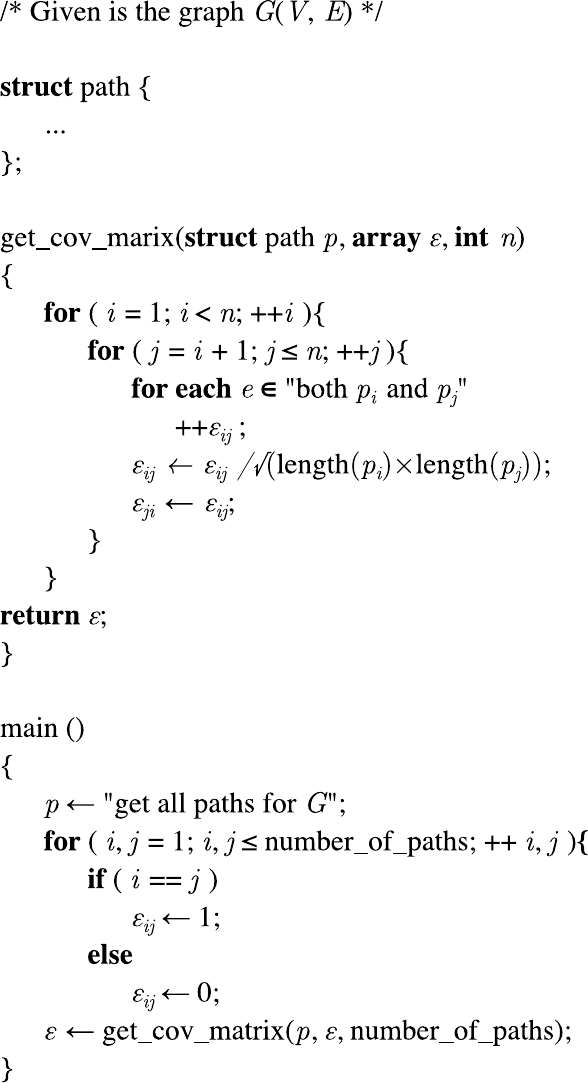}
\caption{A pseudo-code summarising the algorithm for the calculation of the covariance matrix of a given timing graph.}
\label{chap4:fig:pseudo-code_cov_matrix}
\end{figure}

Now, considering the combinations of type $\langle \epsilon_i \epsilon_j \rangle$ and taking into account the relations \eqref{eq:path_2}, the covariance  matrix~\eqref{eq:graph_matrix} can be constructed. Note that it is symmetrical since  $\langle \epsilon_i \epsilon_j \rangle$ = $\langle \epsilon_j \epsilon_i \rangle$. The procedure of the covariance matrix estimation is summarised in the form of an algorithm, and its pseudo-code is given in Fig.~\ref{chap4:fig:pseudo-code_cov_matrix}. The resulting covariance matrix for the graph shown in Fig.~\ref{chap4:fig:graph_covmatrix_algo} can be written as follows:
\begin{equation}\label{eq:graph_matrix}
    \Sigma
    = 
    \begin{pmatrix}
        1 & \frac{3}{2\sqrt{6}} & \frac{3}{2\sqrt{5}} & \frac14 \\
          & 1                   & \frac{4}{\sqrt{30}} & \frac{3}{2\sqrt{6}} \\
          &                     & 1                   & \frac{1}{\sqrt{5}}  \\
          &                     &                     & 1
    \end{pmatrix}\,,
\end{equation}
where the blank spaces imply that the matrix entries are symmetric.

\begin{figure}[!h!]
\centering
\includegraphics[width=0.5\textwidth]{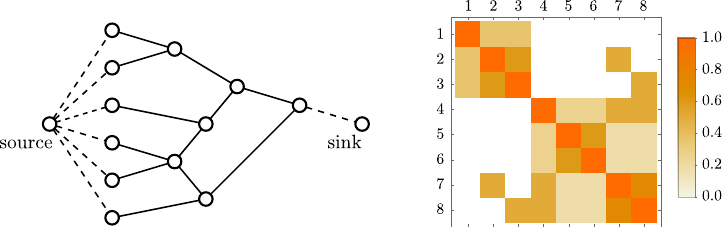}
\caption{Sample timing graph with eight paths and its covariance matrix visualised. This timing graph is used as a `building' block to construct a larger graph shown in Fig.~\ref{chap4:fig:cascade_and_matrixplot2}.}
\vspace{0.2cm}
\label{chap4:fig:graph_and_matrixplot}
\end{figure}

\begin{figure}[!h!]
\centering
\includegraphics[width=0.5\textwidth]{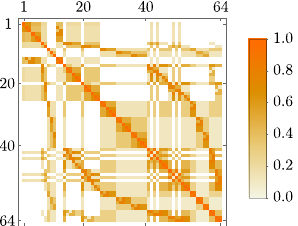}
\caption{Cascade formed by replicating the building block from Fig.~\ref{chap4:fig:graph_and_matrixplot} and its covariance matrix visualised. There are 64 paths in the resulting cascade. }
\label{chap4:fig:cascade_and_matrixplot2}
\end{figure}

To illustrate the algorithm and the result of its operation, we consider a larger timing graph. Firstly, we  construct a ``building'' block, which is a simple timing graph containing several nodes, see Fig.~\ref{chap4:fig:graph_and_matrixplot}. This simple graph is replicated  to construct a larger graph shown in Fig.~\ref{chap4:fig:cascade_and_matrixplot2}. The application of the algorithm shown in Fig.~\ref{chap4:fig:pseudo-code_cov_matrix} yields the covariance matrices, also shown in these figures. The matrices are visualised as a square field and with a colour gradient to convey the magnitude of an entry in a matrix. We note that the entries of a covariance matrix in are $\Sigma_{ij}\in [0,1]$, as negative correlations between paths has no physical meaning.

\section{Discussion and Conclusions}

In this paper, original research contributions are presented. The path-based SSTA problem for VLSI, namely the problem of finding  critical paths in a timing graph, was mapped onto the problem of correlated extremes. For such a problem, approximate solutions in the form of corrections to the Gumbel law~\eqref{chap4:eq:gumbel} for the case of weak correlations were obtained and compared with numerical Monte Carlo simulations. In particular:
\begin{itemize}
    \item An original formulation of the problem was proposed, and the path-based SSTA problem of a timing graph was studied from the perspective of Extreme Value Statistics. This point of view allowed us to obtain  the  upper bound on the worst case delay of VLSI. The upper bound is given by the Gumbel law. 
    \item By obtaining these corrections to the Gumbel law, the mean value of~\eqref{chap4:eq:delay}, which is the total delay in a timing graph, was obtained. The total delay is inversely proportional to the correlation coefficient $\rho$.
    \item A general approach for studying weakly correlated Gaussian RVs, alternative to the known techniques from Refs.~\cite{gyorgyi08,gyorgyi10}, was proposed. The starting point of the approach was the characteristic function~\eqref{eq:characteristic_fun} in the case of $N$ correlated Gaussian RVs. Then, using the relaxation condition~\eqref{eq:smallness_condition} for the covariance matrix $\boldsymbol{\Sigma}$, asymptotic formulae for the CDF of weakly correlated Gaussian RV were obtained by treating non-diagonal entries $\varepsilon_{ij}$ of the covariance matrix as perturbations. The results~\eqref{eq:gumbel_perturbed} and \eqref{eq:gumbel_perturbed_second} extended the Gumbel law~\eqref{chap4:eq:gumbel} in a form of series in powers of the smallness parameter $\varepsilon_{ij}$.
    \item The obtained asymptotic results were in a very good agreement with numerical experiments. Moreover, the second order approximation~\eqref{eq:gumbel_perturbed_second} gave an improvement over the first order one~\eqref{eq:gumbel_perturbed}, as one can see from the PDFs shown in Fig.~\ref{chap4:fig:gumbel_pdf_table}. Studying the behaviour of the  mean  of \eqref{chap4:eq:delay} as the function of the correlation coefficient $\rho$, it was remarkable to obtain even better agreement with the numerical experiments for the cases of the maximum of a relatively small number of RVs, $N=15$, $50$ and $100$. One can see from Fig.~\ref{chap4:fig:gumbel_corr_table} that the second order approximation fits the simulations even in the regime of strong correlations when the theory is not applicable \textit{a priori}.
    \item However, the asymptotic formulae can quickly blow for higher values of the correlation coefficient, when the number of variables in \eqref{chap4:eq:delay} is small, as it is shown in Fig.~\ref{fig:extreme_cdf_pdf}. The influences of the \emph{topology} of the covariation matrix $\Sigma$ on the applicability of the corrections to Gumbel law should be studied rigorously. For this purpose, realistic circuits should be used, which is a scope for a separate study.
    \item The effect of non-identical mean values and variances on the distribution of \eqref{chap4:eq:delay} was studied numerically. It was shown that, when RVs deviate from the identical distributed case slightly, the \emph{average} deviations $\delta\mu$ and $\delta\sigma$, scaled the resulting distribution. This indicated that the proposed theory could be applied for the non-IID case with minimal corrections. The requirement on the deviations $\delta\mu$ and $\delta\sigma$ to be small is justified for logic circuits, as the delays in different paths are not expected to vary greatly. The renormalisation of distributions with respect to $\delta\mu$ and $\delta\sigma$ should be additionally studied in a separate research.
    \item An algorithm for estimation of the covariance matrix $\boldsymbol{\Sigma}$ was proposed. Since it was aimed at obtaining correlations due to shared nodes, the algorithm focused only on the random part of the delays. Assuming the random part $\xi_i$ to be drawn from the standard normal distribution~\eqref{eq:path_delay_random_xi}, the combinations of the average of random components $\langle \epsilon_i \epsilon_j \rangle$ of the path delays were calculated. These combinations corresponded to entries of the covariance matrix.
    
\end{itemize}

\section{Acknowledgement}
The authors wish to thanks Paul Frain, Adrian Wrixon, Anton Belov and all the members of the signoff team at Synopsys, Ireland, for the stimulating discussions.

This work has emanated from research supported in part by Synopsys, Ireland, and a research grant from Science Foundation Ireland (SFI) and is co-funded under the European Regional Development Fund under Grant Number 13/RC/2077.

\appendix

\section{On the Property of Normal Distributions}
\label{app:remarkable_property}


We would like to prove that the expression below is the case:
\begin{equation}\label{app:eq:remarkable_1}
  \left. \frac{\partial \chi}{\partial \varepsilon_{ij}} \right|_{\varepsilon_{ij}=0} = \frac12 \frac{\partial^2 \chi_0}{\partial \mu_i \partial \mu_j}\,.
\end{equation}
To prove this, compute the derivatives directly. For the characteristic function~\eqref{eq:characteristic_fun_factorised}, we have:
\begin{align*}
    \left. \frac{\partial \chi}{\partial \varepsilon_{ij}} \right|_{\varepsilon_{ij}=0}
   & = \left. 
    \left[ - \frac12 k_i k_j \chi_0 (\mathbf k) \cdot \exp \left( - \frac12 \varepsilon_{ij} k_i k_j \right)  \right] \right|_{\varepsilon_{ij}=0}
    \\
    &= - \frac12 k_i k_j \chi_0 (\mathbf k).
\end{align*}

The first two derivatives of the characteristic function of $N$ uncorrelated normal RVs, $\chi_0 (\mathbf k) = \exp \left( i \mu_i k_i  - \frac12 \sigma_i^2 k_i^2\right)$, read
\begin{equation*}
    \frac{\partial \chi_0}{\partial \mu_i}
    =  i k_i \chi_0,
    \quad
    \frac{\partial^2 \chi_0}{\partial \mu_i \partial \mu_j}
    =  - k_i k_j \chi_0.
\end{equation*}

Comparing the left and right sides of the derivatives, we obtain the relation~\eqref{app:eq:remarkable_1}. Calculating higher order derivatives, it is straightforward to show that
\begin{equation}\label{app:eq:remarkable_2}
  \left. \frac{\partial^2 \chi}{\partial \varepsilon_{ij} \partial \varepsilon_{kl}} \right|_{\varepsilon_{ijkl}=0} = \frac14 \frac{\partial^2 \chi_0}{\partial \mu_i \partial \mu_j \partial \mu_k \partial \mu_l}.
\end{equation}

\bibliographystyle{IEEEtran}

\balance

\bibliography{edalib}

\end{document}